\documentclass[12pt]{article}
\begin{document}

\begin{centering}
{\LARGE \bf{Ex-nihilo II: Examination Syllabi and \\ the Sequencing of Cosmology Education}} \\
\ \\ Kevin A.\ Pimbblet \\
Department of Physics, University of Durham, South Road, Durham, 
DH1 3LE, UK.\\ E-mail: K.A.Pimbblet@durham.ac.uk \\
\ \\
John C.\ Newman \\
St.\ Leonard's RCVA Technology College, North End, Durham, DH1 
4NG, UK. \\ \end{centering}
  
\section*{Abstract}
Cosmology education has become an integral part of modern physics courses.  
Directed by National Curricula, major UK examination boards have developed syllabi 
that contain explicit statements about the model of the Big Bang and the strong 
observational evidence that supports it.  This work examines the similarities and 
differences in these specifications, addresses when cosmology could be taught within 
a physics course, what should be included in this teaching and in what sequence it 
should be taught at different levels.

\section{Introduction}
Contained within the frameworks of UK National Curricula, the model of the Big 
Bang is a requisite part of modern day physics teaching (see Pimbblet 2002 for a 
fuller discussion).  For example, the English National Curriculum states that pupils 
should be taught 'about some ideas used to explain the origin and evolution of the 
Universe'.  Building upon these curricula, the major examination boards in the UK 
incorporate statements about Universal origins in their syllabi (see Appendix~A). 
There is, however, little guidance about when to teach cosmology (both within a 
physics course and at what point in schooling), what topics and issues to cover and in 
what order to teach them.  This plan of this work is as follows.  We examine what 
topics are required to be taught, firstly at G.C.S.E.\ level and then at A-level.  Within 
each of these areas, we define the sequence of topics to be taught.  Finally we address 
when cosmology should be taught within any given physics course.

\section{Cosmology in science courses for 14-16 year olds}
All of the G.C.S.E.\ specifications (Appendix~A) require an understanding of the 
Hubble relation.  In some cases this is explicit in the form of $v=Hd$, in others, it is 
implicitly suggested via a qualitative relationship between recession velocity and 
distance (Hubble and Humason 1931).  Given that the Hubble relation represents one 
of the major cornerstones of evidence in favour of an evolving Universe, this is of 
little surprise.

Many of the examination syllabi, however, delve little further into cosmology 
education than Hubble''s relation.  It is of credit to EdExcel that its 
course goes into a 
little more depth.  Firstly, there is the topic of the future evolution of the Universe.  
Depending on the amount of mass and energy that the Universe possesses, one of 
several fates may befall it.  If the Universe has enough matter, then it may cease 
expanding and start to contract back under gravitational force.  This would result in a 
'Big Crunch' scenario.  Conversely, with very little matter contents, the Universe 
would simply go on expanding forever.  The figure of merit that determines which 
fate awaits the Universe is known as the critical density and represents a quantity of 
matter that is just sufficient to cease the Universal expansion.  Modern observations 
display a trend in favour of the latter scenario (Perlmutter et al 1999).  Further, the 
inflationary scenario (e.g.\ Guth 2000) provides a theoretical backdrop for constraining 
the critical density to be very close to one (i.e.\ the Universe just manages to avoid 
collapsing back in on itself).

Related to this topic is the issue of dark matter.  It is thought that much of the matter 
in the Universe has not been (and probably cannot be) observed directly (e.g.\ Peebles 
1993).  Therefore, the so-called cold dark matter (e.g.\ Governato, Ghigna \& Moore 
2001; Colberg et al 2000) will add a significant amount of matter to the content of the 
Universe and hence will influence its future evolution (see above).  

The final topic that appears in some G.C.S.E.\ course specifications is the cosmic 
microwave background radiation, arguably one of the most important astronomical 
discoveries of the twentieth century (Penzias and Wilson 1965).  If interpreted as 
highly redshifted radiation from the Big Bang, it provides unrivalled evidence for an 
evolving Universe that was once extremely hot- several billion Kelvin (see Pimbblet 
2002 for further discussion of this point).

Interestingly, the G.C.S.E.\ EdExcel syllabus also makes explicit reference to the 
'Steady State' theory of the Universe.  It is easy, perhaps, to forget that the Big Bang 
theory was at one time just one of many competing theories (see Ellis 1987 for a 
review of alternative cosmologies).  In the first half of the twentieth century, Hoyle 
and collaborators proposed the rival steady state theory.  In simple 
terms, the steady 
state theory describes a Universe that is homogeneous, isotropic and 
isochronal.  That 
is to say, almost the same as the Big Bang model apart from that it appears identical 
no matter what point in time it is viewed at (i.e.\ has no definite beginning).  Whilst it 
can explain an expanding Universe, steady state predicts that there must 
also be a 
continuous creation of matter: something that has never sat well with the astronomy 
community.  The fall from grace for steady state came with the discovery 
of the 
cosmic microwave background (Penzias and Wilson 1965; see above), which only the 
Big Bang model provides a compelling, natural explanation for.

Therefore, within any G.C.S.E.\ course, we advise teachers to commence cosmology 
with a review of some of the observational evidence in favour of the evolutionary Big 
Bang model: the Hubble relation and the cosmic microwave background radiation.  
This can then readily be underscored with a discussion of the future evolution of the 
Universe.  Finally a whole class discussion about other cosmological theories, 
including steady state, can take place (Pimbblet 2002).

\section{Cosmology in advanced pre-university courses}
The A-level specifications (Appendix~A) broadly follow the same pattern as the 
G.C.S.E.\ ones.  They concentrate on the observational foundations of the evolving 
Big Bang theory (see above) but also touch on other topics.

For example, in the OCR specification is Olbers' paradox.  Named after Wilhelm 
Olbers (1758-1840), the paradox is an old astrophysical issue (see Jaki 1969 for an 
authoritative summary of pre-twentieth century work).  Simply put, the paradox asks 
why the night sky is so dark?  If the Universe is of an infinite age and the stars that it 
contains are distributed evenly (i.e.\ homogeneous and isotropic), it is fairly straight 
forward to conclude that the night sky should be equal in brightness to the Sun (e.g.\ 
Tipler 1988).  Olbers' own resolution to this paradox was to conceive of invisible 
interstellar dust absorbing the light.  Yet, this explanation is insufficient: the amount 
of dust required would obscure the Sun during the day!  Work that followed 
demonstrated that in order for the night sky to appear luminous, the Universe must 
possess an age of $10^{23}$ years.  Therefore, the assumption of an infinite age for the 
Universe is invalid.  Yet, authors also overlooked two important factors for some 
time: stars have finite ages (hence they burn out) and special relativity (hence each 
photon of light that arrives carries less energy than when it was emitted).  Whilst 
Harrison (1987) shows that the dominant factor is the finite ages of stars, both effects 
contribute in the same way: to make the sky darker and thus resolve the paradox.

One major part of cosmology that is conspicuous by its absence from A-level is the 
abundance of the elements that results from nucleosynthesis (e.g.\ Burles, Nollett and 
Turner 2001).  In simple terms, Big Bang nucleosynthesis explains why there is an 
abundance of light elements in comparison to heavier elements.  As such, it provides 
cosmologists with a very good method of testing the quantitative predictions of Big 
Bang theory (Krauss and Romanelli 1990).  We advocate that teachers include 
nucleosynthesis in any advanced level course as, taken in combination with the 
Hubble relation and the cosmic microwave background radiation, they make the Big 
Bang theory appear highly watertight.

Finally, although not on any examination syllabus examined, there are further pieces 
of observational evidence pointing towards an evolving (and hence 
non-steady state) 
Universe.  Such evidence should only be taught to high ability classes when time 
permits.  For example, the Butcher-Oemler effect (Butcher and Oemler 1984) shows a 
recent, strong evolution within the stellar populations of galaxies.  At a diluted level, 
this effect demonstrates that the fraction of blue, star-forming (young) galaxies within 
clusters of galaxies increases with increasing redshift (and hence with decreasing time 
since the Big Bang).  Thus, clusters of galaxies that are further away are less evolved 
and younger than those located nearby.

Therefore, any advanced level course should broadly follow the sequence outlined for 
G.C.S.E.\ courses.  We advise teachers to build upon the observational evidence in 
favour of the Big Bang theory: the Hubble relation, cosmic microwave background 
radiation and include nucleosynthesis.  Olbers paradox can potentially be slotted in 
after this, or at the end of teaching about stellar evolution.  As time permits, other bits 
of evidence such as the Butcher-Oemler effect can also be included as evidence in 
favour of the Big Bang.  The sequence would then follow the G.C.S.E.\ outline again: 
the future evolution of the Universe and a guided class discussion about alternative 
cosmologies.

\section{Sequencing cosmology education}
Having outlined what topics to teach and in which order to teach them in, we now 
turn to the question of when cosmology should be taught within a given physics 
specification.

Astrophysics as a discrete unit of teaching typically comes last within any G.C.S.E.\ or 
A-level scheme of work.  Since the topic requires a synthesis of prior knowledge from 
many parts of a syllabus, this is of little surprise.  The downside is, of course, that 
teaching astrophysics as the last subject will probably not leave sufficient time for it.  
Attempting to teach this topic earlier, say at the beginning of the final year of a 
course, may prove productive, especially given its timeless popularity (e.g.\ Toscano 
2002).  Instead of being a synthesis for other topics, astrophysics can readily be turned 
into a springboard for them.  Thus we advocate teaching astrophysics in the middle of 
a physics course, after some groundwork in classical physics such as forces has been 
taught.

Within astrophysics, cosmology nearly always comes last.  The reason for this 
primarily appears that astrophysics is taught lengthwise as a bottom-up subject: 
starting off with Earth-bound phenomena and working up in scale through the Solar 
System to the Universe as a whole.  The bottom-up method is, however, a sound 
premise because it institutes in pupils a sense of Universal size.

Finally, throughout our discussion about what topics to include in cosmology (see 
above), we have emphasized an observational approach.  This has been done for two 
reasons.  Firstly, any successful cosmological theory (such as the Big Bang) must be 
able to explain the observations.  Secondly, it is predicted that such an observational 
approach will help to deal with many misconceptions that pupils hold about 
cosmology (Pimbblet 2002; Prather, Slater and Offerdahl 2002).

\section{Conclusions}
This work has discussed what sequence cosmology should be taught in (within both 
the 14-16 year old age range and in pre-university courses), what topics to include and 
at what point in schooling it should be taught.

We have suggested that:
\begin{enumerate}
\item Astrophysics as a discrete unit should be taught in the middle of a course once 
sufficient grounding in classical physics (e.g.\ forces) is completed.  It can then 
be used as a springboard into other topics (e.g.\ light).
\item Cosmology should be the last subject within an astrophysics unit.
\item Cosmology education should be built upon the observational foundations that 
support the Big Bang theory (Hubble's relation and the cosmic microwave 
background radiation at G.C.S.E.\ with the addition of nucleosynthesis at A-
level).  Any successful cosmology must, after-all, be able to explain such 
observations.
\item Both Olbers' paradox and the Butcher-Oemler effect broadly support the case 
for an evolving Universe and can be taught as necessary and desired.
\item A discussion about the future evolution of the Universe and other cosmologies 
(Pimbblet 2002) should then follow.
\end{enumerate}

This work follows Pimbblet (2002) and is the second paper in a series examining 
aspects of cosmology education.

\section*{Acknowledgements}
KAP and JCN thank the staff and students of St.Leonard''s RCVA technology college, 
Durham.

\section*{References}
Burles S, Nollett K M and Turner M S 2001 Astrophys. J. 552 L1 \\
Butcher H and Oemler A 1984 Astrophys. J. 285 426 \\
Colberg J M et al 2000 Monthly Not. R. Astron. Soc. 319 209 \\
Ellis G F R 1987 A. Rev. Astr. Astrophys. 22 157 \\
Einstein A 1950 The Principle of Relativity London: Metheun \\
Governato F, Ghigna S and Moore B 2001 in Astrophysical Ages and Times Scales, 
ASP Conference Series 245 469 \\
Guth A H 2000 Phys. Rep. 333 555 \\
Harrison E 1987 Darkness at Night Harvard University Press, Cambridge MA \\
Hubble E and Humason M L 1931 Astrophys. J. 74 43 \\
Jaki S L 1969 The Paradox of Olbers' Paradox Herdr \& Herder, New York\\
Krauss L M and Romanelli P 1990 Astrophys. J. 358 47 \\
Moore G S M 1992 Progress of Theoretical Phys. 87 525 \\
Peebles P J E 1993 Principals of Physical Cosmology Princeton Series in Physics, 
Princeton University Press, NJ \\
Penzias A A and Wilson R W 1965 Astrophys. J. 142 419 \\
Perlmutter S et al 1999 Astrophys. J. 517 586 \\
Pimbblet K A 2002 Phys. Educ. 37 512 \\
Prather E E, Slater T F and Offerdahl E G 2002 Astron. Educ. Rev. Issue 2 (see 
aer.noao.edu) \\
Thomas O 2002 Phys. Educ. 37 492 \\
Tipler F J 1988 Quart. J. R. Astron. Soc. 29 313 \\
Toscano M 2002 Phys. Educ. 37 464 \\

\section*{Appendix~A}
We provide in Table~1 a brief survey of examination syllabi from the major 
examination boards in England, Northern Ireland and Wales.  Scotland has been 
excluded from this survey simply because its examination structure is different from 
that of the other Kingdoms.  Although limited in scope to UK examination boards, the 
content of non-UK physics course specifications, where a statement is made about 
cosmology, are broadly similar in nature.  Readers from outside the UK, however, 
may be surprised at the knowledge expected of students for G.C.S.E.\ level (age 14-
16) and A-level (age 16-19), especially given an already heavily loaded teaching 
schedule (c.f.\ geophysics; Thomas 2002).  Additionally, we note in 
passing that 
cosmology education is typically only given to students who are expected to achieve 
the higher grades (C or above at G.C.S.E.\ level) and is usually not required in 
foundation level G.C.S.E.\ physics courses.
\ \\
\begin{small}
{\bf{Table 1.}}  These are the results from surveying the major 
examination boards' syllabi 
for cosmology education content.  Each syllabus is analysed for content and this is 
presented in the 'categories' column.  A letter 'H' denotes reference to Hubble's 
relation, either implicitly or explicitly; ${\rm{\mu}}$ indicates 
reference to the cosmic 
microwave background radiation; ${\rm{\Omega}}$ indicates reference to the 
future evolution of 
the Universe; 'DM' shows explicit reference to dark matter whilst 'Olbers' denotes 
reference to Olbers' paradox.
\ \\
\begin{tabular}{|l|l|c|}
\hline
Examination board, & Exemplar & Categories \\
type and year. & statement & \\
\hline
CCEA G.C.S.E. & Describe the Big Bang model for & H \\
physics (2004) & the creation of the Universe & \\
\hline
AQA G.C.S.E. & This suggests that the whole Universe & H \\
physics (2003) & is expanding and that it might have & \\
 & started, billions of years ago, from & \\
 & one place with a huge explosion (Big Bang) \\
\hline
EdExcel G.C.S.E. & Describe the Big Bang theory of & H, $\mu$, \\
astronomy (2003) & the origin of the Universe and consider & $\Omega$, DM \\
 & other theories such as the steady state theory & \\
 & Explain how the future evolution of the & \\
 & Universe depends on the amount of mass present. & \\
\hline
OCR G.C.S.E. & Interpret given information about & H, $\Omega$ \\
physics (2003) & developments in ideas on the origin & \\
 & of the Universe & \\
\hline
WJEC G.C.S.E. & Understand that these ideas support & H \\
physics (2003) & a model of an expanding Universe & \\
 & which originated approximately 12 & \\
 & billion years ago with the Big Bang. & \\
\hline
AQA A-Level & (Hubbles law)  Qualitative treatment & H \\
physics (2003) & of Big Bang theory & \\
\hline
OCR A-level & Describe qualitatively the evolution of & H, $\mu$, \\
physics (2003) & the Universe from $0.01s$ after the & $\Omega$, Olbers \\
 & Big Bang to the present & \\
\hline
\end{tabular}

\end{small}
\ \\
Of significant note, the exemplar G.C.S.E.\ statement from AQA suggests that the 
Universe started from one place.  This is a common misconception.  From Einstein's 
field equations of general relativity (e.g.\ Einstein 1950), it is known that 
$G_{\mu \nu} = 8 \pi G c^{-4} T_{\mu \nu}$.  
For a flat space-time, the $G_{\mu \nu}$ components will vanish.  They will 
also vanish for an absence of matter and pressure.  The startling bottom 
line is that 
space-time is generated by matter itself.  Therefore, to say that the 
Universe started 
from one place is simply wrong: with no matter, there could not have been any 
'place', anywhere, to define!  Thus it is gratifying to see that AQA 
have deleted the phrase from one place for their 2004 G.C.S.E.\ syllabus.

\end{document}